\documentclass[preprint, a4paper]{aastex}
\usepackage{natbib}
\shorttitle{Gravitational Mesoscopic Constraints}
\shortauthors{Dantas & Ramos}

\begin{document}

\title {Gravitational Mesoscopic Constraints in \\
Cosmological Dark Matter Halos}

\author{Christine C. Dantas}
\affil{Instituto de Aeron\'autica e Espa\c co (IAE/CTA), \\
P\c ca. Mal. Eduardo Gomes, 50,\\
CEP 12.228-904 - Vila das Ac\'acias \\
S\~ao Jos\'e dos Campos - SP - Brazil\\
{\tt ccdantas@iae.cta.br}}

\and

\author{Fernando M. Ramos}
\affil{Laborat\'orio Associado de Computa\c c\~ao 
e Matem\'atica Aplicada, \\ INPE/MCT, CP 515, \\ 
S. J. dos Campos, 12201-970 SP,  Brazil. \\
{\tt fernando@lac.inpe.br}}

\begin{abstract}
We present an analysis of the behaviour of the `coarse-grained' 
(`mesoscopic') rank partitioning of the mean energy of collections
of particles composing virialized dark matter halos in a $\Lambda$-CDM
cosmological simulation. We find evidence that rank preservation depends
on halo mass, in the sense that more massive halos show more
rank preservation than less massive ones. We find that 
the most massive halos obey Arnold's theorem (on the ordering of the 
characteristic frequencies of the system) more frequently than less massive 
halos. This method may be useful to evaluate the coarse-graining level
(minimum number of particles per energy cell) necessary to reasonably
measure signatures of `mesoscopic' rank orderings in a gravitational
system.
\end{abstract}


\section{Introduction}

Dissipationless N-body simulations of stellar systems indicate that
scaling relations such as the so-called `Fundamental Plane'
(hereon, FP), that is, the
systematic deviation from the expectations of the
virial theorem applied to these systems \citep*[e.g.,][]
{djo87,dre87,pah98,hjo95,bek98,djo88,djo93,dan00}, 
could be reproduced from the final products
of hierarchical merging of galactic model progenitors 
\citep{cap95}. However, not all evolutionary conditions lead to FP-like 
relations: simple gravitational
collapses do not.  That is, objects resulted from mergers form a slightly
non-homologous family (and a FP-like relation), whereas collapses are 
homologous among  themselves (and show no deviation from the virial 
expectations; see \citealt{dan01}).  

At the same time, Kandrup and collaborators  \citep{kan93} argued on the
existence of  `mesoscopic constraints' of pure gravitational origin
in systems relaxing towards virialization  
(hereon, the `Kandrup Effect'). These constraints
were inferred from the general preservation of the 
`coarse-grained' partitioning of the ranked energy distribution of particles, 
and seemed to regulate somehow the  gravitational evolution of these galaxy models 
towards equilibrium.  These constraints were also indirectly
shown to be partially `broken' (violated) in mergers and fully operative
in collapses \citep{dan03}. 

The effect of incomplete mixing of phase space in dissipationless 
gravitational collapses was known already since the decade of 80s
 \citep*[e.g.,][]{whi79,van82,may84,qui88,zar96}. 
The surviving memory of initial conditions in the sense of  an 
almost linear dependence of the final (after the collapse)  
energies with the initial energies (in cosmological  initial conditions)  
was first demonstrated in \citep{vog91}.
A more detailed investigation of this effect in N-body systems 
resulting from cosmological collapses is given in \cite{vog94}.

Such clues lead us to inquire whether the 
`Kandrup Effect' and the scaling relations of gravitational
systems (like the FP)  could be deeply related in some way. 
Here we present a `global map' indicating  where
mesoscopic constraints could be mostly operative, in a full
cosmological simulation. This paper is organized as follows.
In Section 2, we study the `Kandrup Effect' in terms of dark matter halos.
In Section 3, we investigate the behaviour of halos in terms of
Arnold's theorem on the ordering of charcteristic frequencies under
the imposition of a linear constraint. In Section 4, we discuss our
results.

\section{Rank Ordering Preservation of Energy Cells in Dark Matter Halos}

In the study of Kandrup et al., the
distribution of the energy of the particles in systems resulting from
collisions and merging of two model galaxies was analysed in detail.   
They have found that there is a `coarse-grained' sense in which the
{\it ordering} of the mean energy of given collections of particles of
the systems is strictly {\it not violated} through the gravitational
evolution of the models towards equilibrium.  

The method consists of sorting the particles of a  
given initial model according to their
energies. The models are partitioned into a few,
`mesoscopic' (around 5 to 10) bins
of equal number of particles and for each of these bins, the mean
energy is calculated. Finally, the bins are ranked with the first one
initially containing the most bound particles (the most negative mean
energy) whereas the last bin contains the least bounded particles (the least
negative mean energy). The mean energies of these same collections of
particles are then recalculated for the final model and compared with
their initial values. From such an analysis, Kandrup et al.
found that the mean energy rank ordering of fixed collections of
particles is preserved along the evolution.

Here analyse the `Kandrup Effect' in
larger gravi\-tationally-\-dominated structures, like clusters and
superclusters  of galaxies (see also \citealt[]{dan05a}). 
To this end, we have analysed a
$\Lambda$-CDM N-body simulation output of the VIRGO Consortium.  The
analysis is identical to that of \citet[]{dan03}, but here the initial
condition is the z=10 simulation box, and the final condition, the z=0
box (the boxes have a $\sim 250$ Mpc comoving size, where each
particle has a mass of $6.86 \times 10^{10} M_{\odot}$). Signs of the
`Kandrup Effect' were searched for
the $31$  most massive halos found in the z=0 box,
identified by the use of a simple  `friends-of-friends' algorithm
(\citealt{fof}), setting periodic boundary conditions and a searching
length of $1$ Mpc. 

The energy of a particle considered in our work is the mechanical comoving 
one-particle energy. It was not calculated with respect to the local 
center of mass of the particular clumps, but with respect to the comoving
reference frame (that is, the frame which moves with the cosmological
expansion of the simulation box). The comoving energy of a particle $i$ 
was calculated classically from:

\begin{equation}
E_i = 1/2 m v_i^2 - 1/2 \sum_{j \neq i} Gm^2/|{\bf x}_i - {\bf x}_j|,
\end{equation}
with comoving position ${\bf x}_i$ and peculiar velocity $v_i$.  
Units used were Mpc for length, Gyr for time and $M_{\odot}$ for mass.  
The energy associated to the dynamics of expansion of the cosmological 
box does not enter into the above computations.

At this point we remark that in the present simulation scenario ($\Lambda$-CDM), 
the nonlinear collapse of sub-galactic mass halos are the first expected events
after recombination. These small mass units will subsequently cluster together 
in a hierarchy of larger and larger objects (bottom-up structure formation scenario). This is in contrast to top-down pictures where the formation of 
very massive objects comes first, as for instance, in the Hot Dark Matter 
scenario. From the spherical top-hat collapse model, a reasonable estimate for 
the limit to the redshift at which a given halo becames virialized or 
formed ($z_{vir}$) is \citep{lon}: 

\begin{equation}
(1 + z_{vir}) \leq 0.47 \left ( {v \over 100~ km/s} \right )^2 
\left ( {M \over 10^{12} ~M_{\odot}} \right )^{-2/3} \left (
\Omega_0 h^2 \right )^{-1/3}.
\end{equation}
The less massive halo analysed from the set of 31 objects
has a mass of $10^{13} M_{\odot}$. Assuming that 
its velocity dispersion is of order  $\sim  (700 ~km/s)^2$ 
(a typical figure for that mass scale) and the last term of the expression 
above is of order  $\sim 1$, we find that $z_{vir}$ for this halo is 
approximately $\leq 4$. Higher mass halos will have their
$z_{vir}$'s even smaller than in above case. For instance, the most massive halo
in the simulation has $M = 10 ^{14.38} M_{\odot}$. Assuming $v \sim
(10^3 ~km/s)^2$, we find $z_{vir} \leq 0.2$. Since all the 31 halos analysed 
have masses within that range, their condition at $z=10$ (the initial dump of 
the simulation) reasonably represents a linear stage of their evolution. 
Another way to see this point is that the particle mass of the simulation is
$10^{10.84} M_{\odot}$. Hence, such a particle is representing a large 
sub-galactic object, and considering its internal velocity dispersion of $v \sim (300 ~km/s)^2$, then $z_{vir} \leq 24$. Of course, this is the mass resolution of 
the simulation, it is assumed to already represent a collapsed unit at $z=10$. 
In other words, that gives an idea of the redshift formation 
limit for the smallest mass unit resolved in the simulation. 
Hence, the 31 most massive halos at $z=10$ 
(the first redshift dump available from the simulation) 
are safely in their linear phases of evolution, so that the comparison of the
`Kandrup Effect' seems quite satisfactory to be performed in reasonably 
equal grounds for all 31 halos.

We have defined a `violation' parameter rate, $\theta$,
which measures {\it the degree of rank ordering violation  of the mean
energy of $10$ fixed collections of particles}, compared at $z=10$ and
$z=0$, so that if $\theta \rightarrow 0$, we retrieve the `Kandrup
Effect' (no energy rank ordering violation), whereas $\theta
\rightarrow 1$ means that all energy bins presented
ordering violation. Notice that the $\theta$ parameter
``penalizes" cases where there is a significant relative change in energy
cell position (in the sense of ordering). For example, if only the 
10th energy cell crosses all the other $9$ cells, and all these $9$ cells 
keep their ordering unaltered, the $\theta$ parameter in this case is 
{\it not} assigned $0.1$ (meaning that only one energy 
cell, out of $10$, crosses some other cell - whatever how many places 
in ordering), but in fact our criteria assigns $\theta = 1.0$, 
meaning that {\it all} cells are relatively 
crossed in this case. So the $\theta$ parameter, although
not a weighted parameter, measures in fact a relative crossing percentage.
An illustration of the `Kandrup Effect' and its violation, measured by the $\theta$
parameter, is presented in Figures 1a and 1b, for the 31 halos in order
of  increasing mass.

We remark that one expects that the values of 
energies of the particles in a clump (after the formation-collapse of it) 
are spread in a wider range than the range of their initial energies. 
During the collapse and relaxation, a number 
of particles in the clump loose energy and they are trapped in a deeper potential 
well near the center of the clump, while other particles gain energy and they may 
even escape the clump after the collapse. This produces a wider range of energies 
at the final configuration. However, in the various panels of Figs. 1a and 
1b we see that the final energies are almost always distributed in a 
smaller range than the initial energies (with only three or four exceptions). 
Such a description is generally the correct expectation
for energies evaluated at a fixed background. But in the present case, the 
energies are calculated with respect to the comoving frame. 

A qualitative understanding of the results of Figures 1a and 1b is the
following. First, consider two particles ($1$ and $2$) comoving with the expansion. 
Then, after a given time $t^{\prime} > t_0$ (where $t_0$ is the initial
time considered), their comoving coordinates are $x_1(t_0)=
x_1(t^{\prime})$; $x_2(t_0)= x_2(t^{\prime})$, and their potential
energy (with respect to the comoving frame) is unchanged, although they
might seem drifting apart from each other 
for an observer at a fixed reference frame.
Now consider that the two particles just detach from the
expansion (`turn-around'): even though with respect to
a fixed frame their potential energy may look unchanged, it does get
more negative with respect to the comoving frame (because
$x_1(t_0) > x_1(t^{\prime})$; $x_2(t_0) > x_2(t^{\prime})$). 
Subsequently, the particles get bound, their potential energy gets 
more negative in both fixed and comoving frames, but it is then
clear that it gets even more negative in the latter case, because 
the background is expanding. 
The same reasoning is natural to escaping particles. Also, notice
that the velocity used is the peculiar velocity, which is the velocity
substracted from the velocity of expansion. The overall result
to the energy calculated this way is that the particles will present
final energies almost always distributed in a smaller range than the 
initial energies.

Notice that Figs. 1a and 1b refer to the initial and final {\sl mean} energies
of fixed collections of particles. The particles have been initially ranked 
according to their energies (comoving), and binned to a fixed number of particles 
per ranked energy cell. The initial distribution of the energies within 
a given `mesoscopic' energy cell is relatively uniform. However, 
when the halos collapse or merge, this distribution tends to get
skewed towards more negative energy values. Some few particles
do escape and carry energy, forming a tail in the distribution.
The mean energy per cell gets more negative due to the effect explained
above. 

Also, in most cases of Figs. 1a and 1b, the final energies are 
concentrated in lower values than the initial energies, but in 
one case particles seem to have gained energy as they form a clump.
The method used to isolate the cosmological halos is friends-of-friends,
which tends to extract overdensities in a given distribution of
points. We expect that several such overdensities may be considered
as virialized or quasi-virialized halos, but some (few) could in fact 
be artifacts, one of which may be the suspect case mentioned here.

In Figure 2,  $\theta$ is plotted in terms of the mass
of the identified structures in the z=0 box. This Figure shows that  {\it
there is a sense in which larger and larger structures seem to evolve
towards a preservation of the `mesoscopic' mean energies}, whereas
smaller and smaller structures tend to violate the `Kandrup
Effect'. Intermediate-sized structures present  intermediate values of
$\theta$.  

From the above trend, we infer that the dynamics of the smaller
structures is probably being dominated by merging processes, whereas
larger structures seem to be ruled by the collapse mechanism
(intermediate-sized clusters could be collapsing structures accreting
small mass systems). In order to address this question, we attempt to
quantify whether it is mainly the inner radial bins of the halo
which violate mean energy rank, or the outer ones.

At this point, it is important to notice that in a spherical 
gravitational potential, $\Phi(r)$, the orbit of a given particle 
is confined to a plane defined by its angular momentum vector, $\vec{L}$. 
The circular orbits are the ones which have the greatest $L$ by unit mass. 
For a given energy per unit mass, $E$, among all possible $L$s, there will be 
a maximum $L_{circ}(E)$, which corresponds to the circular orbit
(c.f. \cite {bin}). The curve $L_{circ}(E)$ divides the [E,L] 
plane into a forbidden (above the curve) and a populated (below the curve) 
region, were the particles of a given gravitational system will lie. 
(Deviations from spherical symmetry are recognized as the presence of 
a few particles above the $L_{circ}(E)$ curve in the forbidden  region). It can be easily shown that the curve $L_{circ}(E)$ tends asimptotically to $L
\rightarrow 0$ for very bound (negative) particle energies, and $L
\rightarrow \infty$ for $E \rightarrow 0^-$. Hence, particles with
very negative energies (boundest particles) tend to have lower $L$, and
tend to be at the inner radii of the system; on the other hand, particles less gravitationally bound tend to have a larger range of $L$s and mainly populate 
the outer radii of the system. Hence it can be reasonalbly assumed that 
the most negative energy bins analysed are composed of particles mostly 
at the inner radii of the halos.

Having said this, we have analysed the behaviour of the 3 innermost and 3 
outermost energy bins, as explained in the following. For each halo, we have assigned a value of 1 (=YES) whenever rank violation 
is found within each of these two sets of 3 bins (separately), 
and assigned value 0 (=NO) if rank is preserved. Figure 3 illustrates 
the results found. It can be seen that the 3 outermost energy bins violate
rank ordering in 19 out of the 31 halos, whereas for the 3 innermost 
bins the statistics is 29 out of 31. So, whenever rank violation occurs, 
it does mainly at the most bounded energy bins, which means that, 
for the most part, the particles at the inner radii of the model are 
mainly affected, not the outer ones. 

\section{Characteristic Frequencies of Collections of Particles in
Dark Matter Halos}

At this point, two natural questions are:  first, could the
phenomenum observed in Fig. \ref{fa4.5.mer}  be artificaly created
solely because of the different number of particles involved in 
each halo? That is, could it be that, depending on the number of 
particles (namely, mass) of the individual halo in question, the system could 
artifically be more fine-partitioned relatively to a system with more particles, and
hence the former system could show up more violation of the `Kandrup Effect'
than the latter?  In fact, Kandrup et al. have analysed the problem
of coarse-graining (partitioning the energy space into 5 and 20
bins) and found no significant dependence on coarse-graining.
We adopt a cautionary position and identify the halos in which
poor statistics may be a problem. We find that 7 out of the 31 evaluated 
halos have total number of particles 
less than 200, which implies less than 20 particles per bin.
Figure 2 indicates the halos where the phenomenum (namely, rank violation
as mass decreases) could be an artifact 
due to mass (particle) resolution. However, for the majority of
halos, the number of particles per energy cell is reasonably 
larger than the number of coarse-grained cells adopted ($10$ cells, 
in the present case). 

A second, related question is: (i) what is the  coarse-graining level relevant 
to the observation of the `Kandrup Effect', and (ii) why it happens?  
Concerning the latter part of the question, one of the possible 
explanations for the `Kandrup Effect', given by Kandrup et al. themselves, 
but not proven, is the existence of some {\it constraint} operative
only at the level of {\it collections} of particles ranked by mean energy. 
We try to find some clues on this problem by attempting to
answer the first part of the above question. 
One independent method which may prove interesting is to directly verify
the validity of the theorem described by \cite{arn}, concerning
the behaviour of the characteristic frequencies of a dynamical system 
under the imposition of a linear constraint. 

Arnold's theorem describes how the characteristic
frequencies ($\omega$) of a system with $n$ degrees
of freedom are distributed relatively to the characteristic
frequencies ($\omega^{\prime}$) of a system obtained from
the former under the imposition of a linear contraint,
reducing the dimensionality of the given system to $n-1$.
The theorem goes as follows. Let
\begin{equation}
\omega_1 \leq \omega_2 \leq ... \leq \omega_n
\end{equation}
be the $n$ characteristic frequencies of the original
system where no constraints are imposed, and
\begin{equation}
\omega_1^{\prime} \leq \omega_2^{\prime} \leq ... \leq \omega_{n-1}^{\prime}
\end{equation}
the $n-1$ characteristic frequencies of the system with
a constraint. Then the theorem asserts that:
\begin{equation}
\omega_1 \leq \omega_1^{\prime} \leq \omega_2 \leq \omega_2^{\prime} 
\leq ... \leq \omega_{n-1} \leq \omega_{n-1}^{\prime} \leq \omega_n.
\end{equation}

Let us define the characteristic frequency of a particle of the
simulation as $\omega \equiv |\vec{L}|/|R|^2$, where  $\vec{L}$ is the
angular momentum of the particle relatively to the comoving center of mass (CM)
of the system and $R$ the comoving distance of the given particle to the
CM. We have taken the mean frequency values of particles
grouped according to the same rank ordering method of Kandrup et al.,
and then proceeded as follows.
We focused on  the five halos that mostly violate the `Kandrup Effect' 
and the five ones mostly obeying it.
We have partitioned these halos into $9$ orderly energy
coarse grainings (bins) at $z=0$.  At $z=10$, however, the $9$ corresponding
bins were repartitioned into $10$ bins. This allowed us to directly compare
the characteristic mean frequencies of each bin per halo at the two
different redshifts. That is, from Eq.(4) above,
the primed values now refer to the mean frequencies of each of
the $9$ bins analysed at $z=0$ (per halo), and the unprimed values, to the
$10$ bins at $z=10$ (per halo). Then, for each halo, we were able to
analyse whether Arnold's theorem (Eq. 4) would be violated or not, and
at what level. We found that the percentages of halos
matching Arnold's theorem, bin by bin, for the halos mostly obeying
the `Kandrup Effect' (in order of decreasing mass), were: $100 \%$, $100 \%$,
$77.8 \%$, $66.7\%$, and $77.8\%$. For the halos violating energy rank: 
$33.3\%$, $55.5\%$, $66.7\%$, $66.7\%$, and $44.4\%$.
Hence, the most massive halos tend to obey
Arnold's theorem (on the ordering of the characteristic frequencies)
more frequently than less massive halos.

\section{Discussion}

Our general conclusion is drawn from a combination of previous and
present clues on the phenomenum of relaxation.
Collapses are quite different from mergers, despite the
fact that both are the end results of a dynamics based on fluctuations 
on the gravitational potential. These main differences are
at the heart of the connection between the `Kandrup Effect' and 
the FP $\times$ virial relations. 

From the current interpretation, merging of stellar systems occurs 
due to a transfer of the orbital energy to the particles within 
the systems through tidal interactions. This mechanism increases 
the internal energy of the systems at the expense of their orbital energy.
If the orbital energy is less negative (approaches $0^-$),  
there is plenty of time for the particles  to withdraw
energy from the orbit of the pair of merging models.  This process
involves periodically, slowly evolving, large fluctuations on the
potential, which takes a larger amount of time to stabilize. 
It was already evident that such slow of stabilization process
could be responsable for producing non-homology among 
the simulated mergers \citep{cap95,dan03}. 
Collapses, on the contrary, are generally  much more
`violent' than mergers, even those resulting from relatively `warm' 
initial conditions. 

The present analysis corroborates the idea that collapses and mergers
behave differently from the point of view of their energy rank 
preservation/violation, considering that both mechanisms
are dominant at different scaling (mass) regimes. Indeed, we find that:

\begin{itemize}
\item{More massive halos tend to better
preserve their mean energy rank ordering than less massive halos.}
\item{Whenever rank violation occurs, 
it does mainly at the most bounded energy bins, which means that, 
for the most part, the particles at the inner radii of the model are 
mainly affected, not the outer ones. This is in agreement with the 
qualitative idea that the central region of the halo is where the 
gravitational potential fluctuates more strongly during the collapse, 
whereas in the outer regions of the halo the fluctuations are softer 
and dump rapidly. In the case of mergers, however, it is expected that 
the outer regions will be submitted to equally large fluctuations because of 
the anisotropy of the merging geometry as compared to a simple, isotropic 
collapse. Indeed it is interesting to notice that rank violation of the 3 
outermost bins does occur much more frequently for less massive halos, 
where merging is expected to occur more frequently as well. }
\item{The process of mesoscopic `mixing' in the energy space seems to be 
inefficient as compared with `mixing' in configuration and/or velocity space, 
an effect which is at odds with Lynden-Bell's theory of complete `violent 
relaxation' \citep{lyn67}; see discussion in Kandrup et al. However, this effect
seems to be more evident depending on the process of virialization
(collapses seem more prone to energy rank preservation).}
\item{We provide a general method to evaluate the coarse-graining level
(number of particles per energy cell) relevant to reasonably
measure signatures of the `Kandrup Effect'. We have observed that
Arnold's theorem, applied to the collections of particles of each
energy bin, tends to be valid more frequently when the `Kandrup Effect'
is operative, than when it is not. This may be related to the
presence of a `mesoscopic constraint' operative at the level of particles,
as observed by Kandrup et al. Alternatively, when Arnold's theorem
is shown not to be valid, the `Kandrup Effect' is expected to be
violated. In this case, one could suspect of small particle number 
statistics or, otherwise, intrinsic (real) energy rank violation.}
\end{itemize}

We briefly assess how our results could be affected by a different 
choice of evaluation of energies. Kandrup et al. analysed the merging 
of two galaxies and have computed the energy of a given particle 
considering two choices:

\begin{enumerate}
\item {relative to the common center of mass (CM) of both galaxies;}
\item {relative to center of mass of the galaxy in which the particle
was originally located.}
\end{enumerate}
Kandrup et al. show that, if the initial center of mass velocity
is relatively small, the tendency (preservation of the ordering of
the mean ranked energies of collections of particles) is little 
affected (methods 1 and 2 give similar results), although the effect 
does get weaker if the velocity of the CM  is reasonably large. 
Notice however that the simulations of Kandrup et al. were performed
on a fixed background. In the present situation the 
energies are all comoving quantities,
the velocity of expansion of the cosmological background 
does not enter into the above computations, so it is our opinion
that our choice of evalutation of energies is reasonable in the
cosmological set. A posteriori, our results (Fig.1a and 1b) do 
reproduce the `Kandrup Effect'.

We also consider the question whether our results depend on the position 
of the particular clump inside the simulation box. 
A qualitative reasoning indicates that they must not, 
since we calculate the comoving energies
with no reference to the relative location within the simulation box.
Also, the halos analysed in the paper are distributed at several
locations within the simulation box, and the results are reasonably 
similar among them, irrespective of their position, 
depending mostly on the halo mass.

We would like to draw attention to the relation of our results to
those of the recent paper by \cite{han}. These authors have studied 
the velocity distribution function (VDF) of different gravitational
systems (from isotropic to highly non-isotropic structures), 
simulated from diverse N-body experiments, including cosmological ones,
and found that the VDF has a universal profile. This may indicate that
some fraction of the particles of a gravitational system 
exchange energy during the relaxation process only until the universal 
VDF has been reached. It would be highly interesting to
understand the relation between energy rank preservation and the
universal VDF.

In summary, the scenario presented here, although admittedly schematic, does offer
some preliminary important clues on the origin of the scaling relations 
of virialized systems.
However, a rigorous understanding of the relaxation process
still lacking  \citep{pad90}, in order to reach a complete understanding
of the phenomena outlined in this paper.

\clearpage
{\bf Acknowledgements}

The simulations in this paper were carried out by the Virgo 
Supercomputing Consortium using computers based at Computing Centre 
of the Max-Planck Society in Garching and at the Edinburgh Parallel 
Computing Centre. The data are publicly available at 
{\tt www.mpa-garching.mpg.de/NumCos}.
We thank Dr. Andr\'e L. B. Ribeiro for several discussions, 
and the anonymous referee for useful suggestions.
C.C.D. thanks Instituto Nacional de Pesquisas Espaciais (INPE),
Divis\~ao de Astrof\'{\i}sica (DAS/CEA), Brazil, for using
its facilities when most part of this paper was written.

\begin{figure}
 \includegraphics[width=17cm,height=19cm]{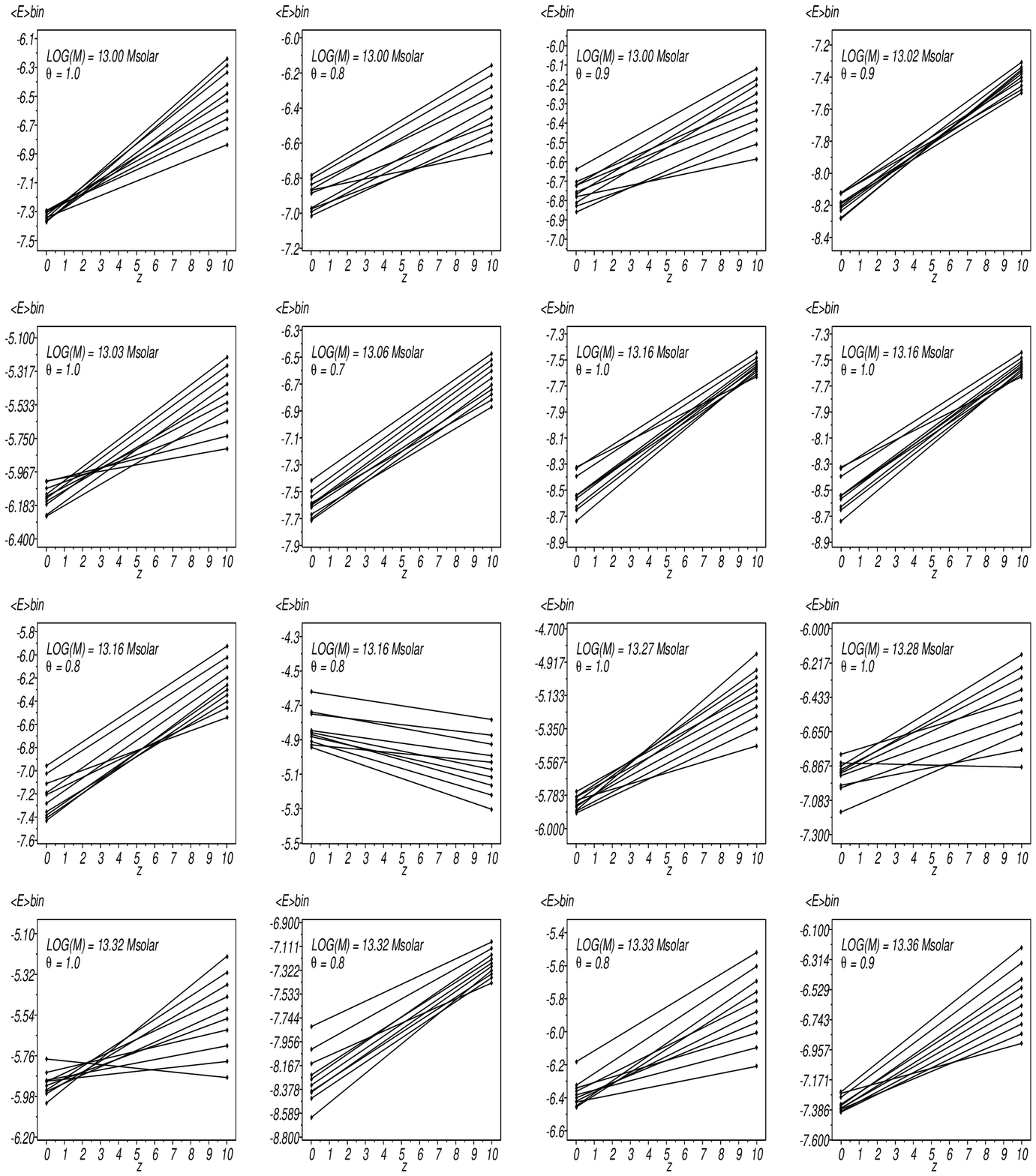}
 \caption{(a)The `mesoscopic' mean energy of fixed
number of particles ($<E>_{bin}$) at $z=0$ (final condition) 
and $z=10$ (initial condition), connected by line segments
in order to illustrate the ordering of the mean energy of given
collections of particles, from the most gravitationally
bounded ones (most negative energies) to the less bounded ones (less
negative energies). The panels are ordered from the less massive halos
(top left panel) to the most massive halos (bottom right panel).
The second panel (from left to right) at the third row is possibly
an overdensity artifact (not a virialized halo).
\label{kan}}
 \end{figure}

\clearpage
\setcounter{figure}{0}
\begin{figure}
 \includegraphics[width=17cm,height=19cm]{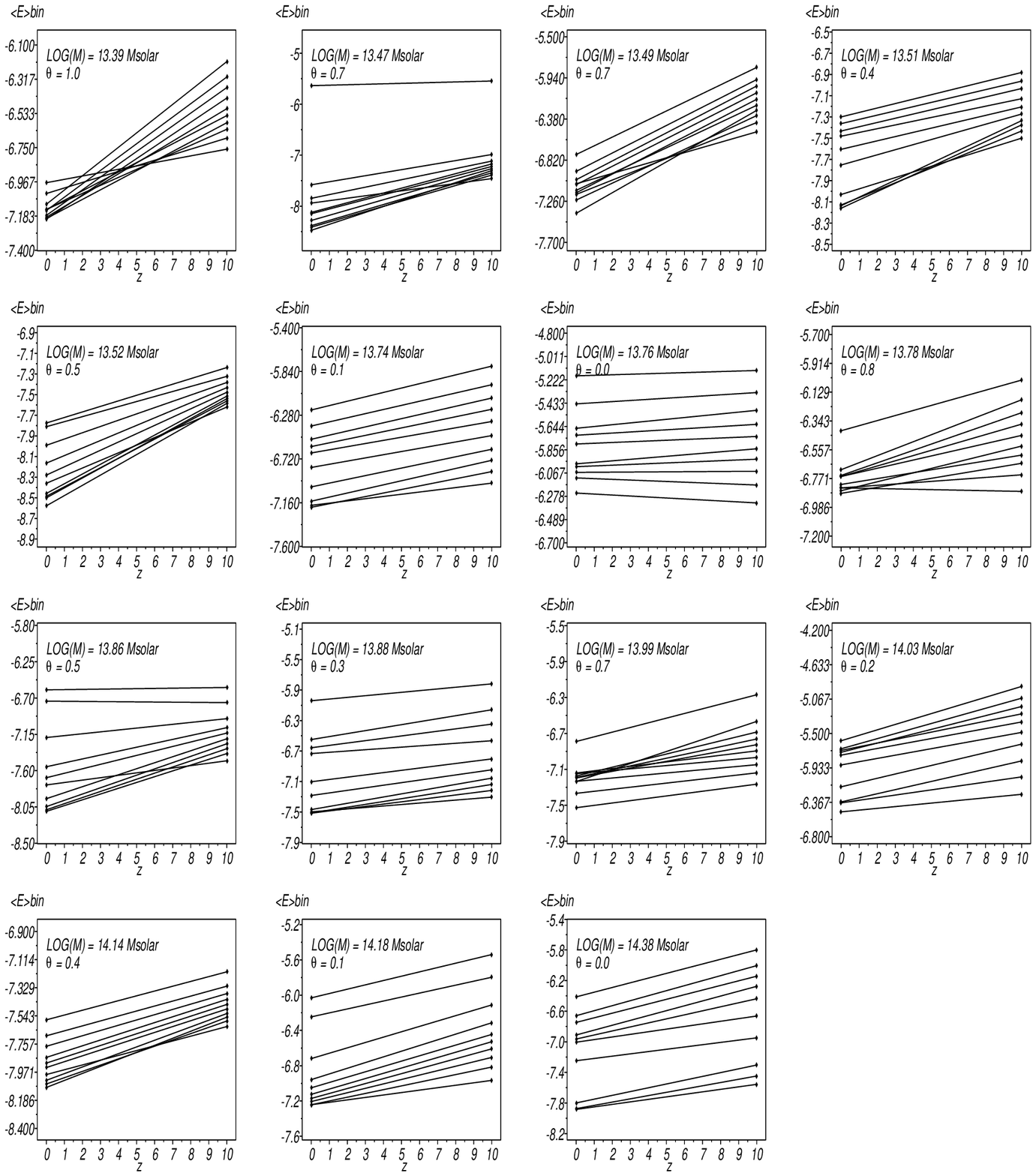}
 \caption{(b) Continued from Fig. 1a.
\label{kanb}}
 \end{figure}

\clearpage
\begin{figure}
 \includegraphics[width=16cm,height=12cm]{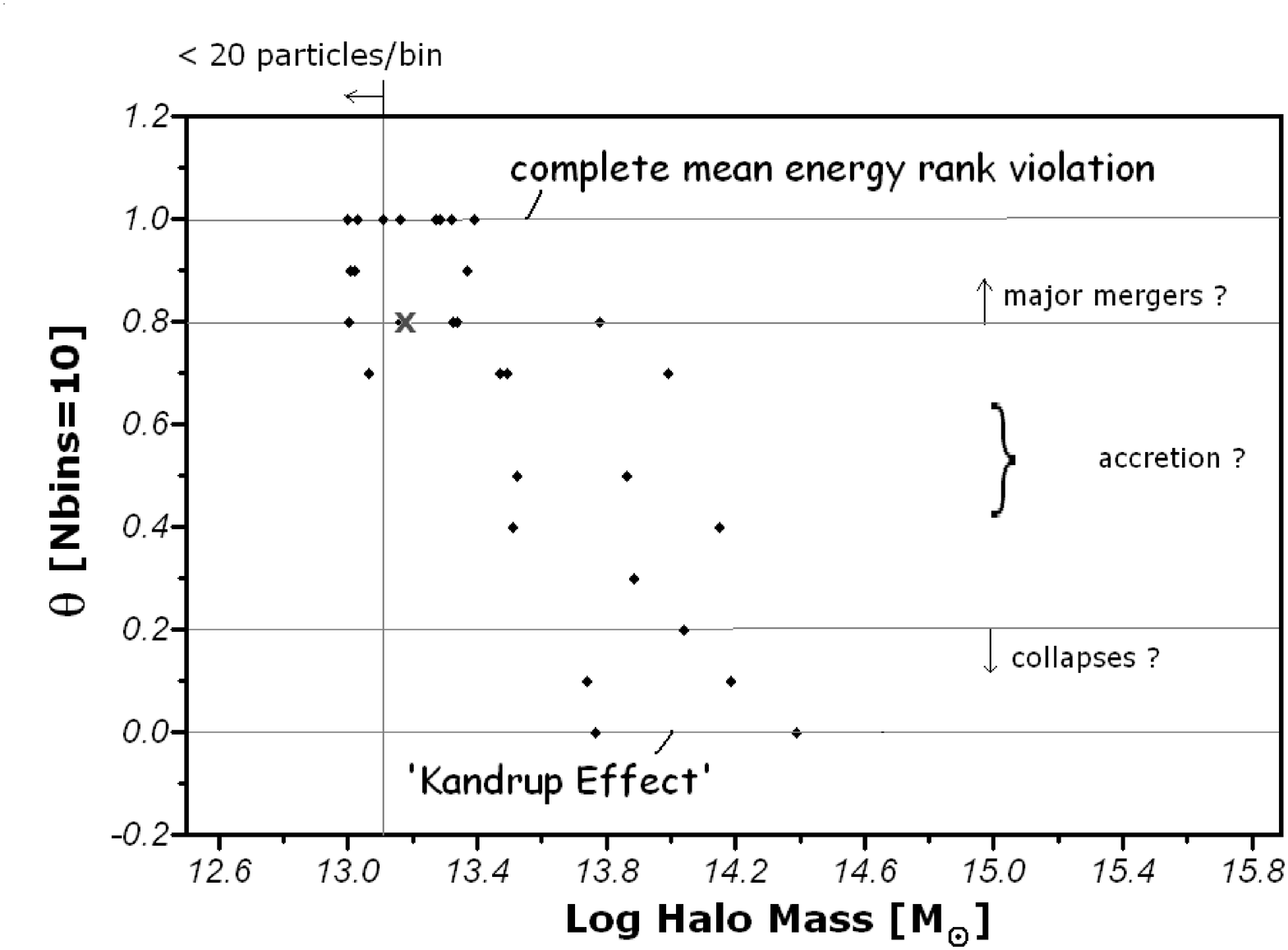}%
 \caption{The `mesoscopic' mean energy ordering violation rate ($\theta$)
as a function of the halo mass in a z=0 $\Lambda$-CDM
cosmological box. Possible dynamical events are identified in this
figure. The object marked with an `X' is possibly an overdensity
artifact (not a virialized halo).
\label{fa4.5.mer}}
 \end{figure}

\clearpage
\begin{figure}
 \includegraphics[width=16cm,height=12cm]{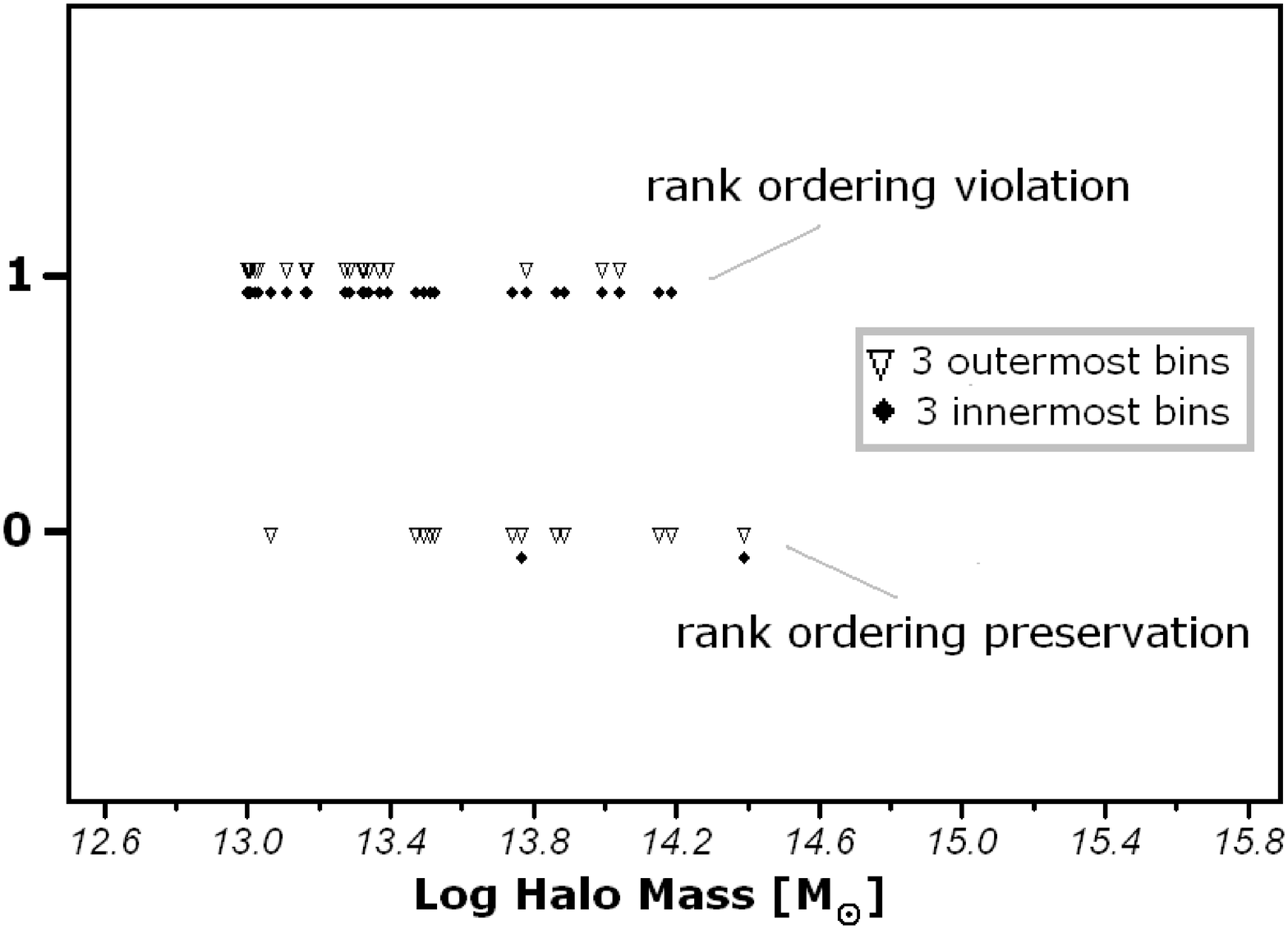}%
 \caption{Analysis of the 3 innermost and 3 outermost energy bins of
all 31 halos considered.
 Violation (preservation) of energy rank is signaled at the vertical
 axis by the value 1 (0).
\label{innerouter}}
 \end{figure}


\begin{thebibliography}{99}
\bibitem [Arnold (1989)]{arn} Arnold, V. I., {\it Mathematical Methods of
Classical Mechanics}, 2nd. edition, 1989, Springer-Verlag, New York.
\bibitem [Bekki (1998)]{bek98} Bekki, K., 1998, ApJ, 496, 713.
\bibitem [Binney \& Tremaine (1988)]{bin} Binney \& Tremaine, Galactic Dynamics,
Princeton University Press.
\bibitem [Capelato et al. (1995)]{cap95}
Capelato, H. V., de Carvalho, R. R. \& Carlberg, R. G., 1995, ApJ,
451, 525.
\bibitem [Dantas et al. (2000)]{dan00}  Dantas, C. C., Ribeiro, A. L. B., Capelato, H. V., 
and de Carvalho, R. R., 2000, ApJ, 528, L5.
\bibitem [Dantas et al. (2002)]{dan01} Dantas, C. C.,
Capelato, H. V., de Carvalho, R. R., \&  Ribeiro, A. L. B., 2002,
A\&A, 384, 772.
\bibitem [Dantas et al. (2003)]{dan03} Dantas, C. C., Capelato, H. V., Ribeiro, A. L. B. \&
de Carvalho, R. R., 2003, MNRAS, 340, 398.
\bibitem [Dantas \& Ramos (2005)]{dan05a} Dantas, C. C. \& Ramos, F. M., 2005, ApJ, submitted.
\bibitem [Djorgovski \& Davis (1987)]{djo87} Djorgovski, S. G. \& Davis,
M., ApJ, 1987, 313, 59.
\bibitem [Djorgovski (1988)]{djo88} Djorgovski, S. G., in Proc. Moriond
Astrophysics Workshop, 1988, Starbursts and Galaxy Evolution, ed. T. X.
Thuan et al. (Gif sur Yvette: Editions Fronti\`eres), 549.
\bibitem [Djorgovski (1992)]{djo92} Djorgovski, S. G., in Cosmology and Large-Scale
Structure in the Universe, ASP Conference Series, Vo. 24, 1992,
Reinaldo R. de Carvalho, (ed.), 19.
\bibitem [Djorgovski (1993)]{djo93} Djogorvski, S. G. \& Santiago, B. X., 
1993, in {\it Structure,
Dynamics and Chemical Evolution of Elliptical Galaxies}, 
Proc. ESO/EIPC Workshop, ed. I. J. Danziger, W. W. Zeilinger, \& K. Kjar
(ESO Conf. and Workshop Proc. 45) (Garching:ESO), 59
\bibitem [Dressler et al. (1987)]{dre87} Dressler, A.,
Lynden-Bell, D., Burstein, D., Davies, R. L., Faber, S. M.,
Terlevich, R. J., \& Wegner, G., 1987, ApJ, 313, 42.
\bibitem [FOF homepage (2005)]{fof} FOF Algorithm, http://www-hpcc.astro.washington.edu/tools/fof.html.
\bibitem [Hansen et al. (2005)]{han} Hansen, S. H. et al., astro-ph/0505420
\bibitem [Hjorth \& Madsen (1995)]{hjo95} Hjorth, J. \& Madsen, J., 1995, ApJ 445, 55.
\bibitem [Kandrup et al. (1993)]{kan93} Kadrup, H. E., Mahon,
M. E., \& Smith, H., 1993, A\&A 271, 440.
\bibitem [Longair (2000)]{lon} Longair, M., Galaxy Formation, Springer,
page 366.
\bibitem [Lynden-Bell (1967)]{lyn67} Lynden-Bell, D., 1967, MNRAS 136 101.
\bibitem [May \& van Albada (1984)]{may84} May \& van Albada, 1984, MNRAS, 209, 15
\bibitem [Padmanabhan (1990)]{pad90} Padmanabhan, T., 1990,
Physics Reports, 188, No.5, 285-362.
\bibitem [Pahre (1998)]{pah98} Pahre, M., 1998, PhD Thesis, California Institute of Technology.
\bibitem [Quinn \& Zurek (1988)]{qui88} Quinn \& Zurek, 1988, ApJ, 331 ,1
\bibitem [van Albada (1982)]{van82} van Albada 1982, MNRAS, 201, 939 
\bibitem [Voglis, Hiotelis \& Hoeflich (1991)]{vog91} Voglis, Hiotelis \& Hoeflich, 1991, A\& A, 249, 5
\bibitem [Voglis, Hiotelis \& Harsoula (1994)]{vog94} Voglis, Hiotelis \& Harsoula, 1994, Astrophysics and Space Science, 226, 213. 
\bibitem [Zaroubi et al. (1996)]{zar96} Zaroubi et al., 1996, ApJ, 457, 50
\bibitem [White (1979)]{whi79} White, S., 1979, MNRAS, 189, 831

\end{thebibliography}
\end{document}